\documentclass[preprint,11pt]{elsarticle}
\usepackage{amsmath}
\usepackage{lineno}
\usepackage{amssymb}
\usepackage[numbers]{natbib}
\usepackage{graphicx}
\usepackage{hyperref}
\usepackage[T1]{fontenc}    
\usepackage[utf8]{inputenc} 
\usepackage{lmodern}        
\usepackage[polish,english]{babel} 
\usepackage{subcaption}
\usepackage[dvipsnames]{xcolor}
\begin{document}


\begin{frontmatter}

\title{Energetic Feasibility of Redirecting Trans-Neptunian Objects onto Mars-Impacting Orbits: Continuous Thrust and Gravity Assist Trajectories}

\author[cbk]{Ryszard Gabryszewski}
\author[cbk]{Leszek Czechowski}
\author[uw]{Arkadiusz Hess}

\address[cbk]{Space Research Centre Polish Academy of Sciences, Warszawa, Poland}
\address[uw]{University of Warsaw, Warszawa, Poland}

\makeatletter
\g@addto@macro\elsaddress{
  \par\vspace{2em}
  \vbox{
    \centering\normalfont\small
    \textbf{Authors' preprint version}\\[0.5em]
This is the authors' preprint of the manuscript, available under the CC-BY-NC-ND 4.0 license: 
https://creativecommons.org/licenses/by-nc-nd/4.0/ \\[0.5em]
    The final published version has appeared in \textit{Icarus}:\\[0.5em]
    \textit{Icarus}, 2026, vol. 457, 117164.\\
    DOI: \url{https://doi.org/10.1016/j.icarus.2026.117164}\par
  }
  \vspace{1em}
}

\makeatletter
\def\ps@pprintTitle{%
  \let\@oddhead\@empty
  \let\@evenhead\@empty
  \let\@oddfoot\@empty
  \let\@evenfoot\@empty
}
\makeatother

\begin{abstract}
We assess the dynamical feasibility of redirecting small volatile-bearing trans-Neptunian objects (TNOs) onto Mars-impacting orbits using continuous low-thrust propulsion and a single gravity-assist encounter. The study considers two representative dynamical classes: classical Kuiper Belt--like and Scattered Disk--like initial orbits, and determines the minimum characteristic velocity increment $\Delta V$ required to drive the objects onto a Mars-impacting trajectory within a specified transfer time $\Delta T$. The dynamics is modelled in the two-body problem with a fixed maximum low thrust included, allowing the computed $\Delta V$ to represent a dynamical lower bound independent of specific propulsion-technical implementation.

Three trajectory classes are investigated: (i) inward spiral transfer, (ii) time-dependent thrust-direction steering optimized via global evolutionary algorithms, and (iii) hybrid transfers combining low thrust with a single Neptune flyby. Pure spiral trajectories yield very high velocity expenditures ($\Delta V \gtrsim 22~\mathrm{km~s^{-1}}$) and millennia durations, confirming that monotonic inward migration is dynamically inefficient for TNO redirection. In contrast, optimized steering strategies systematically increase orbital eccentricity and achieve Mars-impacting geometries with $\Delta V \approx 2.5$--$3.2~\mathrm{km~s^{-1}}$ over 380--540 yr timescales. A single Neptune encounter further reduces the total $\Delta V$ in favourable cases, with minimum values falling below those of direct optimized transfers. These results establish a quantitative lower bound on the energy cost of importing volatiles from the outer Solar System to Mars, showing that controlled redirection is feasible under modest $\Delta V$ budgets when target bodies are chosen from favourable regions of orbital phase space.
\end{abstract}

\begin{keyword}
terraforming \sep Mars \sep Kuiper Belt \sep trans-Neptunian objects \sep orbital mechanics \sep low-thrust trajectories \sep gravity assist \sep Mars impacts \sep orbital dynamics \sep trajectory optimization
\end{keyword}

\end{frontmatter}

  \vspace{1em}

\section{Introduction}

Terraforming scenarios for Mars generally require increasing the atmospheric mass several times relative to the present value of  $\sim 2.5 \times 10^{16}$~kg (e.g., \citet{Czechowski2025}). Even the modest target
pressures associated with the Armstrong limit or pure-oxygen breathing conditions imply
importing $10^{17}$--$10^{18}$~kg of volatile material. The mass requirements are especially seen in the radical version 6 of \citet{Czechowski2025}. They serve as a main motivation for the present work. We consider here
whether bodies from outer Solar System, characterized by abundant volatiles, can be transported to Mars with the
technology, that is possible to reach in a few dozen of years.

The Kuiper Belt represents the substantial reservoir of volatile-rich bodies. The density of a few kilometre-sized objects is typically low, e.g. for comet 67P/Churyumov–Gerasimenko $\rho \approx 0.4$~g\,cm$^{-3}$. These bodies retain abundant H$_2$O, CO$_2$, CO, NH$_3$ and other ices \citep{Paetzold2016, Filacchione2019, Rubin2020}. The total mass of the Belt ($\sim 1/30\,M_\oplus$) is sufficient for any conceivable atmospheric enhancement scenario on Mars.

The dynamical challenge is severe. A body must transition from heliocentric distances of 30--100~AU to $\sim 1.5$~AU, crossing a vast orbital-energy gap. Impulsive propulsion is incapable to accomplish this task for kilometre-scale natural bodies. Continuous thrust, hypothetically achievable through future high specific impulse energy sources \citep{Choueiri2009, Glover2011}, can only deliver minor accelerations, but over century-long operation it can set a proper geometry for an encounter with Mars. The additional use of gravity assists offers an efficient mechanism for reducing the heliocentric energy thereby lowering the total $\Delta V$ required for impact, assuming that favorable encounter geometry is available \citep{Czechowski1991}. 

The idea of employing large-scale engineering interventions to render Mars more habitable has a long intellectual history, with one of the first quantitative treatments was provided by \citet{ZubrinMcKay1993}. Their study introduced a mathematical model of the Martian CO$_2$ system, explicitly coupling the atmosphere, polar caps, and regolith reservoirs. They demonstrated that the greenhouse feedback could reduce the required engineering input by two orders of magnitude compared to earlier estimates made by \citet{McKay1991}, who assumed that sustaining habitable conditions would require the continuous industrial-scale production of artificial greenhouse gases on Mars.

Instead of assuming that all greenhouse gases must be imported, they argued that modest initial forcing - delivered via orbital mirrors, imported volatiles, or \emph{in situ} production of halocarbons - could trigger the release of CO$_2$ and produce atmospheric pressures of several hundred millibars on relatively short timescales. Their computational framework used semi-empirical equilibrium relations for CO$_2$ vapor pressure and regolith adsorption.

Zubrin also investigated the feasibility of delivering volatiles by importing icy bodies from the outer Solar System. His astrodynamical analysis was deliberately simplified: orbital transfers were modeled as Hohmann-like maneuvers, supplemented by gravity assists from the major planets, with the resulting $\Delta V$ estimated in the range of 300~m\,s$^{-1}$ for an object redirected from outside Jupiter's orbit onto a Mars-crossing trajectory. The calculations assumed nearly circular and coplanar orbits, neglecting eccentricity and inclination influence, and considered the asteroid's own volatile inventory (e.g., NH$_3$) as propellant for nuclear-thermal rocket engines. With a specific impulse of $\sim 400$~s, Zubrin concluded that propellant mass fractions of $\sim 10\%$ would suffice to redirect a body of order $10^{10}$ tonnes. He estimated that four such impacts could initiate a greenhouse sufficient to warm Mars. While highly approximate by modern orbital mechanics standards, these estimates were among the first to link planetary-scale climate engineering with dynamical transport of outer Solar System bodies, thereby opening a line of inquiry that subsequent work has sought to refine.

Subsequent analyses have expanded and refined these ideas. \citet{Czechowski2025} emphasized the enormous energetic cost of raising Mars' pressure even to one-tenth Earth's, showing that Kuiper Belt objects provide the most plausible volatile source but are dynamically fragile when transported inward. \citet{Palka2022} employed spatial data science and global climate modeling to quantify the atmospheric energy budget of terraforming, explicitly simulating the impact of cometary nuclei as an alternative to greenhouse gas release. Recent geomorphological and remote-sensing studies further highlight the coupling between subsurface processes and atmospheric evolution \citep{Czechowski2023,Czechowski2025remote}. Concepts for controlled off-world habitats \citep{Adler2020} also illustrate how terraforming arguments interface with broader planetary engineering frameworks.

What emerges is a clear methodological gradient. Early works such as Zubrin's relied on thermodynamic and simplified orbital models, where $\Delta V$ was treated as an order-of-magnitude scaling. A radical proposal - variant 6 in \citet{Czechowski2025} - assumes the transport of a larger mass of volatiles than other versions. Therefore, he proposes a more powerful propulsion system based on thermonuclear reactions (fusion reactions). Thanks to the greater mass of the transported volatiles, the climate will be more stable. It may not require artificial methods of maintaining conditions (e.g., orbital mirrors or factories of greenhouse gases). Its stability should be ensured by living organisms, similarly to Earth. Moreover \citet{Czechowski2025,Czechowski2025remote} consider gravity assists between bodies in the Kuiper Belt, as well as tidal forces (see e.g., \citet{Czechowski1991}) during transport of bodies in Kuiper Belt. Other papers such as \citet{Palka2022} integrate planetary climate models with spatially explicit simulations, coupling impact scenarios with atmospheric dynamics. This evolution illustrates how the field has moved from feasibility arguments toward increasingly data-driven and system-level approaches.

Against this backdrop, we revisit the transport problem from a strictly dynamical perspective, focusing on the minimum control acceleration required to place volatile-bearing trans-Neptunian bodies on Mars-impacting orbits. 
We deliberately step beyond classical impulsive templates such as Hohmann and bi-elliptic transfers, well characterized since the analysis of \citet{HS61}, so that our results are not restricted to circular, coplanar boundary conditions or three-impulse constructs.

While the controlled redirection of small Solar System bodies has long been discussed conceptually, the practical means of achieving it remain beyond current technological capability. Conventional impulsive propulsion, suitable for spacecraft, cannot be applied to kilometer-sized natural bodies due to their large inertia and lack of structural integrity. At the same time, no existing system is capable of continuously imparting even a weak, sustained acceleration to such objects. Nevertheless, future advances in \emph{in situ} resource utilization, laser ablation or other methods may enable long-duration thrusting or outgassing control. For the purposes of this study, we adopt a representative upper bound of $4\times10^{-10}\,\mathrm{km\,s^{-2}}$ for the continuous acceleration. This value reflects the order of magnitude expected from future ISRU-based momentum–transfer systems, large-scale ablation, or related methods acting on kilometre-scale bodies. Although this acceleration is $10^{4}$--$10^{6}$ times smaller than that of conventional rockets, when applied over century-long durations it can accumulate to velocity changes making controlled dynamical transport theoretically feasible.

Recent preliminary work has investigated related transfer pathways using combined continuous thrust control and simulated Neptune encounters \citep{HessCzechowskiGabryszewski2025}, highlighting the need for more systematic optimization approach. In this paper, we extend the dynamical aspect of the problem by analysing the feasibility of redirecting small volatile-rich bodies from two classical Kuiper Belt and Scattered Disk regions onto collision trajectories with Mars. 

Our numerical model employs the two-body problem, neglecting planetary perturbations, in order to determine the minimum characteristic velocity increment ($\Delta V$) and the minimum transfer time ($\Delta T$) required to achieve an impact in each scenario. The equations of motion are integrated under a small continuous control acceleration, interpreted here as a generic dynamical control term rather than the performance of a specific propulsion technology. Throughout the paper, this idealized control acceleration is referred to simply as continuous thrust. We also examine the energetic and temporal benefits of employing a gravitational assist at Neptune. Unlike previous treatments based on Hohmann \citep{Zhang2020} or bi-elliptic transfers which are valid only for circular and coplanar orbits, this study shows how orbital geometry, particularly high eccentricity, can drastically reduce both the energetic cost and the duration of Mars-delivery missions.

\section{Dynamical Model}

In this work, we consider a heliocentric restricted two-body problem in which both Mars and a small test particle of negligible mass move under the gravitational influence of the Sun. Planetary perturbations are omitted so that the resulting trajectories represent the minimum-energy baseline for redirection. In addition to this central force, the particle is subjected to a continuous thrust acceleration. The model 
is used only to explore the dynamical feasibility of orbital redirection. The $\Delta V$ values reported in this work refer only to the orbital redirection phase and do not include the cost of reaching or operating at the target trans-Neptunian object.

In this two-body model, the continuous acceleration vector at the initial time $t_{0}$ remains constant. This assumption eliminates gravitational perturbations and thereby suppresses any time variability of the optimal control law. In a realistic \(N\)-body environment, however, planetary perturbations would continuously alter the orbit of the body, requiring thrust-direction adjustments to preserve the transfer trajectory. Consequently, the \(\Delta V\) budget obtained in the present two-body model represents a minimum, whereas practical implementations within a full \(N\)-body model would inevitably demand larger control effort due to the necessary time-dependent steering. 

\subsection{Equations of motion and control formulation}

The heliocentric two-body problem can be defined as:
\begin{equation}
\ddot{\mathbf{r}} = -\frac{\mu_\odot}{r^{3}}\,\mathbf{r} + \mathbf{f}_{\mathrm{th}},
\end{equation}
where \(\mu_\odot\) is the solar gravitational parameter and \(\mathbf{f}_{\mathrm{th}}\) denotes a constant acceleration (continuous thrust) vector:
\begin{equation}
\mathbf{f}_{\mathrm{th}} = f_{0}\,\hat{\mathbf{u}}(t), \qquad |\hat{\mathbf{u}}|=1.
\end{equation}
A uniform thrust value of:
\begin{equation}
f_{0} = 4\times10^{-10}\ \mathrm{km\,s^{-2}}\;=\;0.0004~\mathrm{mm\,s^{-2}}
\end{equation}
was adopted as the maximum acceleration value in all simulations, representing a propulsion acting continuously on kilometre-sized bodies.

The adopted thrust model corresponds to an ideal continuous acceleration approximation commonly used in preliminary trajectory studies. The magnitude of the acceleration is assumed constant and its direction is controlled according to the selected guidance strategy. This simplified formulation neglects propellant consumption and engineering constraints on the thrust direction in order to focus on the dynamical aspects of the problem.

Numerical integration was performed using the eighth-order Dormand–Prince DOP853 scheme with adaptive step-size control. At each integration step, the contribution to the cumulative control effort was recorded as:
\begin{equation}
\Delta V = \sum_{k} f\,\Delta t_{k},
\end{equation}
providing an exact time-integrated measure of the total velocity increment associated solely with the non-gravitational acceleration. The instantaneous thrust magnitude is constrained by $f \leq f_{0}$.

\subsection{Transfer scenarios}

Three transfer strategies were analyzed:

\paragraph{(a) Spiral transfer} The continuous thrust direction is permanently aligned opposite to the velocity vector producing a gradual logarithmic inward spiral. 
\begin{equation}
\hat{\mathbf{u}}(t) = -\frac{\mathbf{v}}{|\mathbf{v}|}
\end{equation}

\paragraph{(b) Optimal control transfer with thrust-direction steering}
The thrust direction is treated as a control variable selected to minimize the spatial distance from Mars after a transfer time~$\Delta T$: 

\begin{equation}
J = \big\|\mathbf{r}_{\mathrm{ast}}(t_0+\Delta T)
       - \mathbf{r}_{\mathrm{Mars}}(t_0+\Delta T)\big\|,
\end{equation}

The final distance from Mars is used here as a penalty-type objective function. 
This formulation allows the optimization algorithm to search for trajectories approaching the planet without explicitly imposing a hard boundary constraint. When the objective function approaches zero, the resulting trajectory corresponds to a Mars-impact solution. In practice, the optimization therefore seeks trajectories that minimize the terminal distance from Mars, and candidate impact solutions are identified when this distance becomes sufficiently small.

The vector components are determined using global evolutionary optimization methods: the Covariance Matrix Adaptation Evolution Strategy (CMA-ES) and Differential Evolution (DE), with population sizes adjusted to the considered
scenario and a termination threshold corresponding to a positional tolerance on the order of kilometers at impact. The resulting trajectories display time-dependent thrust geometries involving combined tangential, transverse, and normal components. Optimization increased the orbital eccentricity to values exceeding 0.95, effectively elongating the trajectory and leading to minimal thrust acceleration needed for finding the impact with Mars. Consequently, the overall $\Delta V$ requirement became significantly reduced for initial orbits located in the Kuiper-belt region.

\paragraph{(c) Gravity-assist transfers} 

To investigate whether gravitational encounters can reduce the control effort required for Mars-delivery trajectories, we examined hybrid transfers in which continuous thrust is combined with a single close passage near Neptune. In these trajectories, the gravitational encounter is used to reorient the heliocentric velocity vector, thereby lowering the perihelion and shifting the orbit into the vicinity of Mars’ heliocentric distance. The continuous thrust acceleration is applied only to guide the test body into the desired encounter geometry and then to impact Mars, while the primary change in orbital orientation is supplied by the planetary flyby itself.

A single Neptune flyby was modeled in the three-body approximation using the \texttt{Poliastro} library \citet{Cano2023}. 
The deflection was applied as a rotation of the incoming heliocentric velocity \(\mathbf{v}_{\mathrm{in}}\):
\begin{align}
\delta &= 2\,\arctan\!\left(\frac{G M_N}{b\,v_{\mathrm{rel}}^{2}}\right), \\
\mathbf{v}_{\mathrm{out}} &= \mathbf{R}(\delta, \phi)\,\mathbf{v}_{\mathrm{in}},
\end{align}
where $\delta$ is the deflection angle, \(v_{\mathrm{rel}}\) is the hyperbolic speed relative to the planet, \(\phi\) is the azimuthal angle in the \(b\)-plane, and \(GM_{N}\) is the standard gravitational parameter of Neptune.

Both the pre- and post-flyby trajectory segments were numerically integrated, with continuous thrust acceleration included. The total velocity expenditure was defined as:
\begin{equation}
\Delta V_{\mathrm{tot}} = \Delta V_1 + \Delta V_2,
\end{equation}
with \(\Delta V_1\) and \(\Delta V_2\) representing the cumulative control effort in pre- and post-flyby segments, respectively.  

In practice, accurate targeting of the Neptune encounter would require trajectory correction maneuvers during the approach phase. However, such corrections are standard in interplanetary missions and do not significantly affect the order-of-magnitude $\Delta V$ estimates presented here.

\subsection{Initial conditions and integration parameters}

We chose two types of trans-Neptunian orbits to characterize distinct dynamical regimes:
\begin{itemize}
  \item TB \#1: \(a_0 = 45~\mathrm{AU}, e_0 = 0.2, i_0 = 10^\circ\),
  \item TB \#2: \(a_0 = 80~\mathrm{AU}, e_0 = 0.5, i_0 = 35^\circ\),
\end{itemize}
with all other orbital angles set to zero at epoch \(t_0\). The first orbit is typical for classical Kuiper Belt bodies while the second one is typical for Scattered Disk objects. Such orbits were selected to assess how variations in eccentricity and inclination affect both transfer time and $\Delta V$ including the dynamical effectiveness of the Neptune gravity assist in steering the post-encounter orbit toward Mars.

The maximum propagation time was selected to guarantee sufficient dynamical evolution for Mars-impacting opportunities in each scenario: up to 3~kyr for the spiral transfer and 500~yr for the controlled and gravity-assist cases. The adopted methodology therefore combines direct numerical integration of the two-body equations with continuous thrust, explicit accumulation of control effort, and stochastic global optimization of the thrust geometry. 

\section{Results}

The outcomes presented below focus on a representative subset of solutions illustrating the efficiency gains achievable through thrust optimization and planetary encounters in comparison to a spiral transfer reference case. 

\subsection{Spiral transfer}

In the baseline scenario the continuous thrust was applied opposite to the instantaneous orbital velocity, yielding a classical spiral-descent trajectory for both test bodies.

\begin{figure*} 
  \centering
  \includegraphics[width=0.95\textwidth]{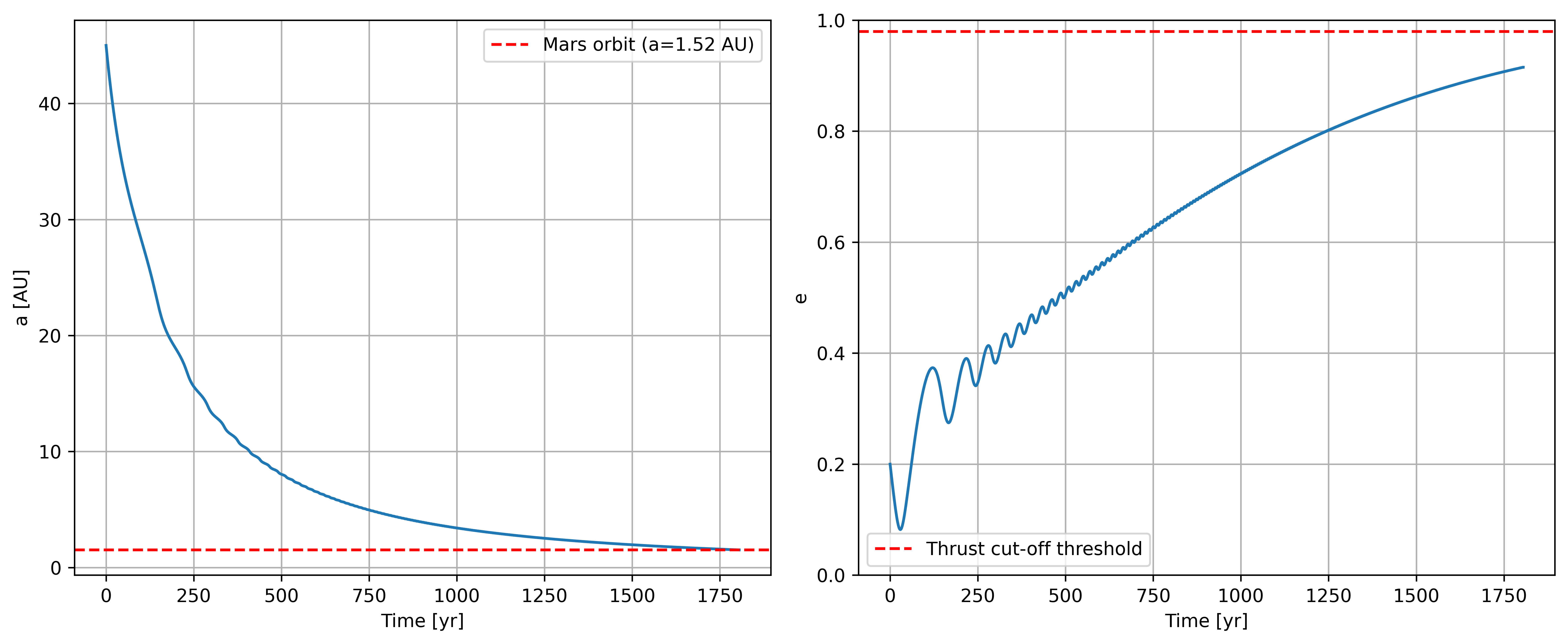}
  \caption{Continuous thrust spiral transfer of the test body~TB~\#1. The panels show the time evolution of the semi-major axis and eccentricity. Thrust acts in the anti-velocity (retrograde) direction along the orbit.}
  \label{fig:1}
\end{figure*}

The time evolution of the orbital elements, shown in the Fig. 1 panels reveals a nearly exponential decrease in the semi-major axis accompanied by a gradual increase in the eccentricity. The contraction of $a$ proceeds in a step-like manner mainly in the intermediate phase of the evolution, while $e$ exhibits a superimposed oscillatory behaviour before entering a secular growth regime. 

The step-like pattern visible in $a(t)$ results from the non-uniform orbital response to the continuous thrust. Although the thrust magnitude remains constant in time and its direction is always opposite to the instantaneous velocity vector, the efficiency of orbital energy removal varies along the orbit. The strongest reduction in $a$ occurs near perihelion, where the velocity - and therefore the thrust-induced work per unit time - is maximal. Between perihelion passages, the change in $a$ is smaller, producing the characteristic “staircase” profile, in which each step corresponds to one orbital revolution with a progressively shorter period as the orbit contracts. This step-like pattern becomes more visible for larger starting $e$. 

The early oscillatory behaviour of $e(t)$ reflects the dynamical coupling between the semi-major axis and eccentricity under a non-central perturbation. Initially, the antitangential thrust component causes alternating phases of increasing and decreasing $e$ depending on the orbital geometry and the orientation of the thrust. As $a$ decreases and the orbital period shortens, the oscillation amplitude is gradually damped due to the growing dominance of the secular component of the thrust. 

Once the orbit approaches the inner Solar System, the evolution becomes predominantly monotonic, with $e$ steadily increasing because the net, orbit-averaged effect of retrograde thrust reduces the specific orbital energy more efficiently than the orbital angular momentum, thereby driving a secular growth in eccentricity.

\begin{figure*}[htbp]
  \centering
  \includegraphics[width=0.95\textwidth]{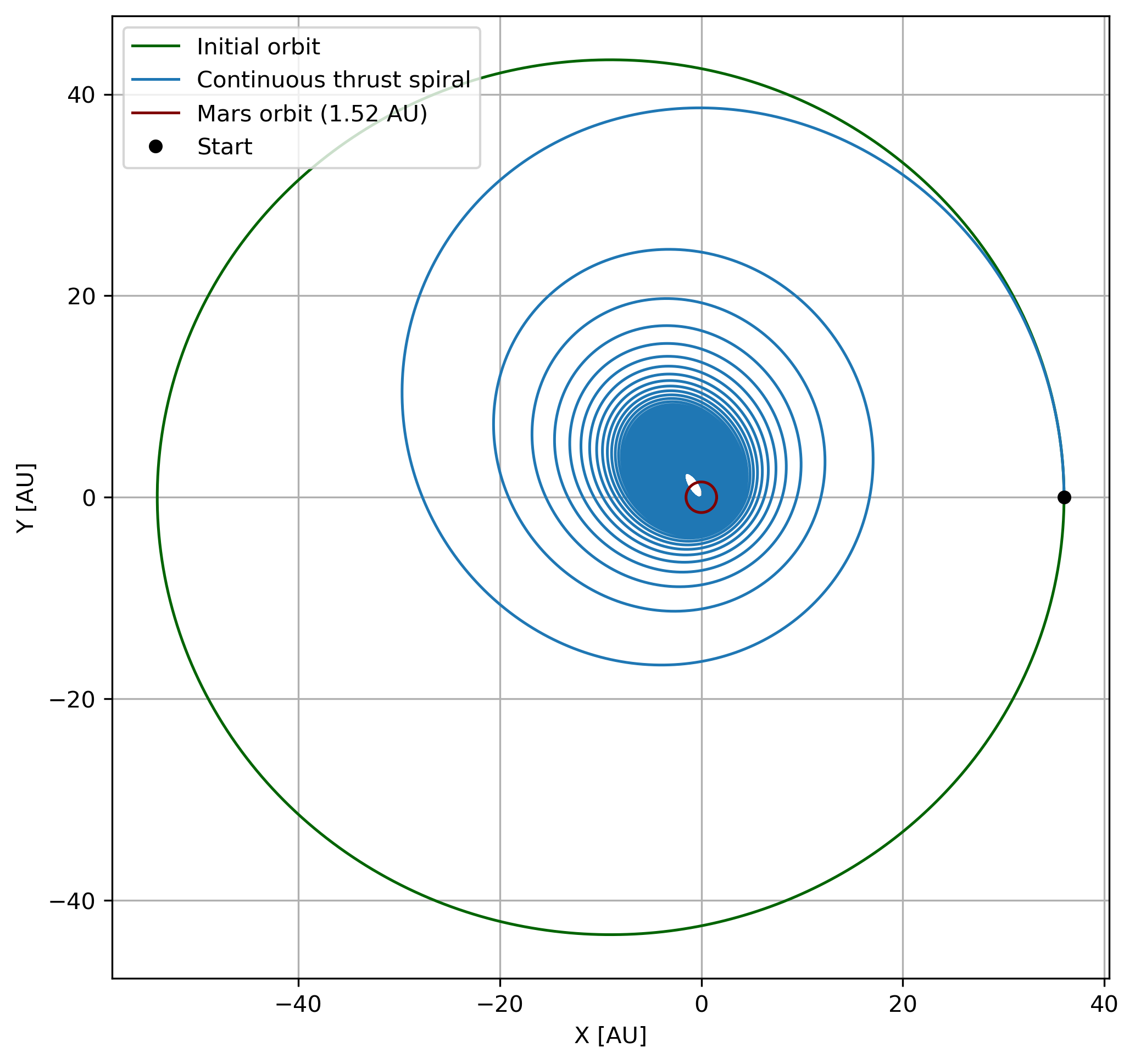}
  \caption{Continuous thrust spiral trajectory in the XY plane projection - TB~\#1 test body.}
  \label{fig:2}
\end{figure*}

Trajectory projected onto the orbital plane on Fig. 2, exhibits the characteristic inward spiral typical of retrograde acceleration. Each loop corresponds to one orbital revolution, with the radius gradually decreasing as the object loses orbital energy. The spiral remains planar, as the thrust acts within the orbital plane, preventing inclination changes. The loops progressively tighten as the trajectory approaches $a_{\mathrm{Mars}} = 1.52~\mathrm{AU}$, because the average heliocentric velocity decreases as the orbit shrinks and becomes more elongated. As the orbit becomes elongated, the average orbital velocity decreases, since the body spends an increasing portion of each revolution near aphelion, where the heliocentric velocity is lowest. This, in turn, reduces the fractional energy loss per orbit and thereby lowers the efficiency of semi-major axis decay.

For TB \#1 approaching the Martian orbital radius, this transfer mode requires a cumulative velocity increment and a time of:
\[
\Delta V = 22.8~\mathrm{km\,s^{-1}}, \qquad \Delta T = 1.805\times10^{3}~\mathrm{yr},
\]
which makes this purely decelerative continuous thrust transfer dynamically inefficient and energetically impractical for trans-Neptunian objects redirection.

In the case of test body~TB~\#2, the retrograde acceleration does not allow the trajectory to reach the Martian orbit, since the eccentricity exceeds $e > 0.999$ before $a_{\mathrm{Mars}} = 1.52~\mathrm{AU}$ is attained (only elliptical orbits are considered in this study). To achieve convergence toward the Martian semi-major axis, we introduced simple modulation of the thrust magnitude similar in the form:
\[
g(r) = \left(\frac{r_{\mathrm{peri}}}{r}\right)^{n},
\]
where $r_{\mathrm{peri}}$ denotes the perihelion distance of the osculating orbit, $r$ is the current heliocentric distance, and $n$ is an exponent controlling how strongly the thrust magnitude decreases away from perihelion. This function reduces the effective thrust away from perihelion, thereby limiting the excessive growth of eccentricity and helping to maintain a more regular inward spiral.

However, this adaptive continuous thrust profile significantly extends the transfer duration to $\Delta T \simeq 2.9\times10^{3}~\mathrm{yr}$ while decreasing the cumulative velocity increment by roughly $10\%$. This shift in thrust utilisation therefore exchanges longer transfer times for improved orbital circularization, highlighting the fundamental trade-off between dynamical control, energetic expenditure, and transfer duration in continuous thrust trajectories.

It should be noted that the examples discussed above do not represent Mars-impacting solutions. Achieving an actual collision would require both orbital intersection and precise phase alignment between the asteroid and Mars, which would further increase the overall transfer time and the cumulative $\Delta V$.

\subsection{Optimized continuous thrust transfer to impact Mars}

\begin{figure*}[htbp]
  \centering
  \includegraphics[width=0.95\textwidth]{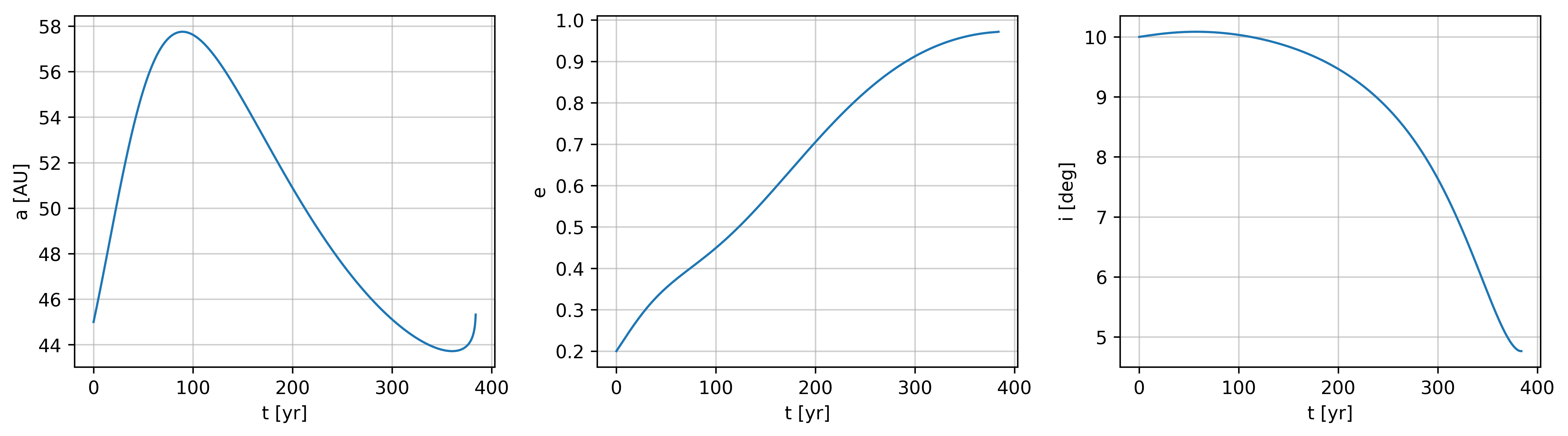}
\caption{Time evolution of the TB \#1 test body orbital elements during the optimized continuous thrust transfer toward a direct collision with Mars. The three panels show variations of semi-major axis, eccentricity and inclination in time which allow the body to reach the collision configuration with minimal propellant consumption.}
  \label{fig:3A}
\end{figure*}

In this scenario, the continuous thrust steering was treated as a time-dependent control variable optimized to produce a direct collision with Mars after an adopted transfer duration $\Delta T$. The optimization primarily acts to reshape the orbit with semi-major axis oscillations between its starting value and  $58~\mathrm{AU}$ (see Fig.~\ref{fig:3A}). The thrust vector efficiently pumps orbital eccentricity by acting near aphelion, where its influence is largest. The monotonic growth of eccentricity, which increases from the initial $e=0.20$ to nearly $e=0.98$, pulls the perihelion distance downward until it coincides with the heliocentric radius of Mars, enabling impact without requiring a large decrease in orbital energy (see the oscillation of the semi-major axis). Thus, the transfer is governed by shaping the orbit into an extremely elongated ellipse.

The inclination evolution (right panel of the Fig.~\ref{fig:3A}) follows the same slow and gradual pattern: it decreases from about $10^\circ$ to below $5^\circ$ over the full duration of the transfer. This adjustment is necessary to match the heliocentric orbital plane of Mars closely enough to allow an actual physical crossing of the trajectories. Because only weak normal thrust component is available, the inclination changes occur on century timescales and constitute one of the limiting factors governing how long the optimized thrust must remain active.

\begin{figure*}[htbp]
  \centering
  \includegraphics[width=0.95\textwidth]{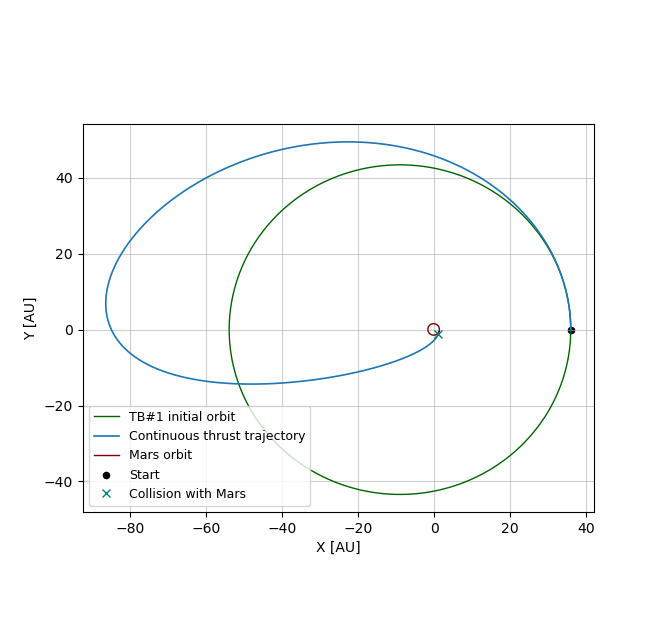}
  \caption{Optimized heliocentric trajectory of TB \#1 under continuous thrust. The trajectory evolves from the initial circular orbit (green) to a highly eccentric transfer trajectory (blue) that intersects Mars's orbit (red) at the designated impact point. The continuous thrust maneuver gradually increases the orbital eccentricity while adjusting 
the apse line orientation to achieve the required collision geometry.}
  \label{fig:3B}
\end{figure*}

The geometry of the resulting trajectory is illustrated in Fig.~\ref{fig:3B}. Once the eccentricity surpasses $e\gtrsim0.9$, the heliocentric motion becomes dominated by long aphelion arcs at large distances ($\gtrsim 40~\mathrm{AU}$) and short, sharply curved perihelion passages near Mars. The optimized thrust predominantly acts during the aphelion portion, where the slow orbital motion maximizes the cumulative effect of weak accelerations on the argument of perihelion, longitude of ascending node, and semi--major axis. The final impact occurs before the first perihelion passage. 

The resulting $\Delta V$ requirements show a systematic dependence on the initial orbital geometry. Test body TB\,\#2, which begins on a more eccentric orbit, requires substantially less total $\Delta V$ than TB\,\#1. This reflects the dynamical advantage of starting closer to the high-eccentricity regime: once the eccentricity already exceeds $e\gtrsim0.5$, continuous thrust induces larger geometric changes in perihelion distance and apsidal orientation. Conversely, test bodies with initially more circular or only moderately eccentric orbits must first be driven into the high-eccentricity domain before efficient collision targeting becomes possible, increasing both the transfer time and the integrated $\Delta V$.

\begin{figure*}[htbp]
  \centering
  \includegraphics[width=0.95\textwidth]{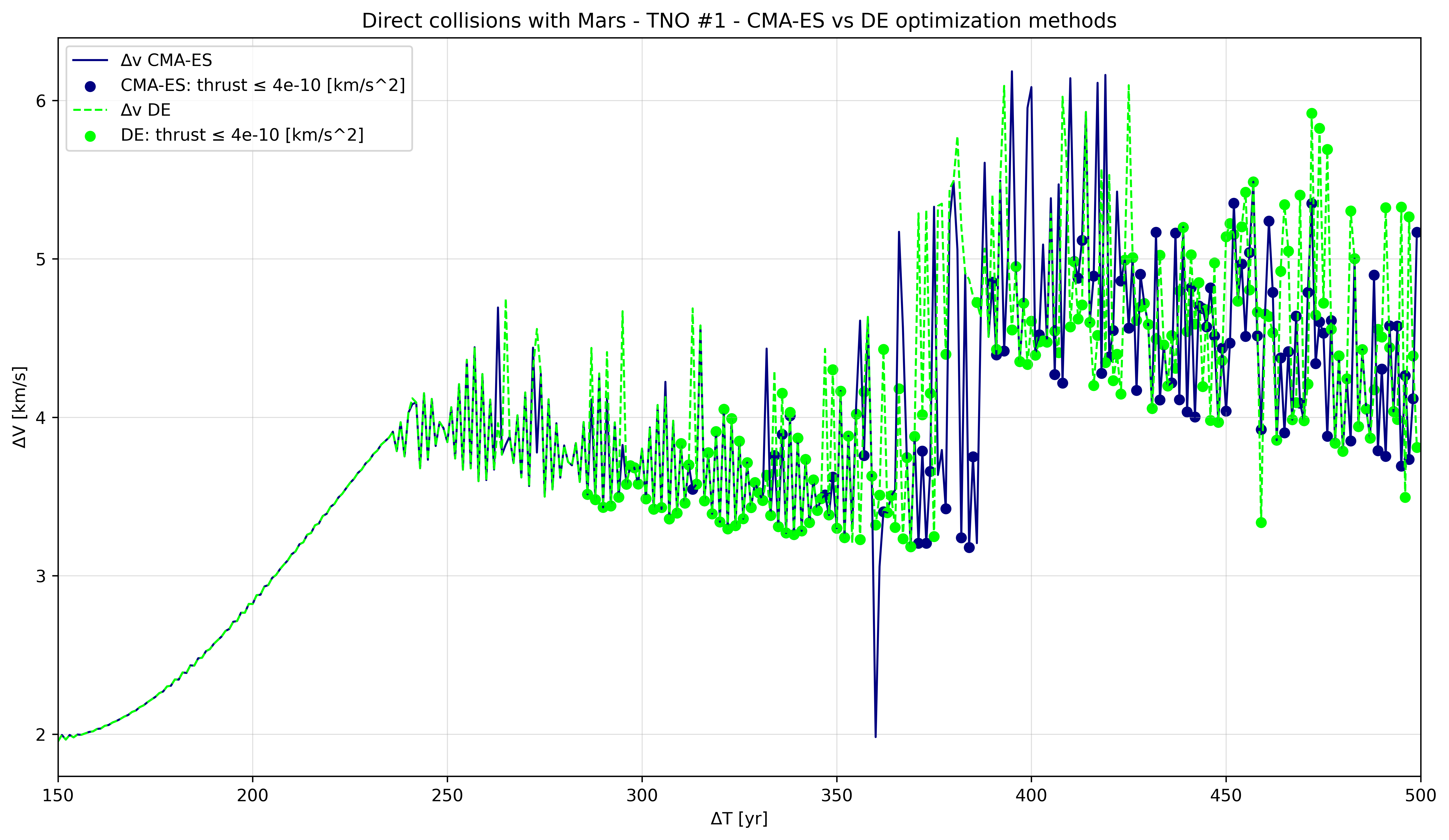}
  \caption{Optimized characteristic velocity increments $\Delta V$ as a function of transfer times $\Delta T$ for direct collision trajectories for test body TB~\#1 with Mars. Solid blue lines indicate solutions obtained using the CMA-ES optimization scheme, while dashed green lines and markers correspond to Differential Evolution (DE) results. The solid circles denote solutions where impacts with Mars are possible within or below the imposed limit of $4\times10^{-10}~\mathrm{km\,s^{-2}}$. Both optimization methods yield similar families of collision-enabling trajectories, with closely matching $\Delta V(\Delta T)$ trends over most of the investigated interval. A noticeable reduction of feasible impacts near $\Delta T \approx 379$--$386$~yr is due to geometry mismatch between the test body orbit and Mars orbit.}
  \label{fig:3}
\end{figure*}

The transfer optimization was perform using two global evolutionary algorithms: Covariance Matrix Adaptation Evolution Strategy (CMA-ES) and Differential Evolution (DE). The $\Delta V(\Delta T)$ curves are presented in Fig.~\ref{fig:3}. 

For transfer time up to about 350~yr, the $\Delta V$ curves obtained with CMA-ES and DE remain broadly consistent, showing only minor deviations in the locations of local minima. For longer transfer times, the results diverge more noticeably: both methods exhibit increasingly irregular oscillations and distinct minima, indicating a more complex and sensitive optimization landscape in which small variations in trajectory parameters lead to substantially different collision geometries with Mars.

There is also a collisional gap at $\Delta T \approx 379$--$386$~yr which corresponds to a phase mismatch between the object and Mars. When the geometry becomes unfavorable, the required thrust effort to maintain collision feasibility increases.
 
\begin{figure*}[htbp]
  \centering
  \includegraphics[width=0.95\textwidth]{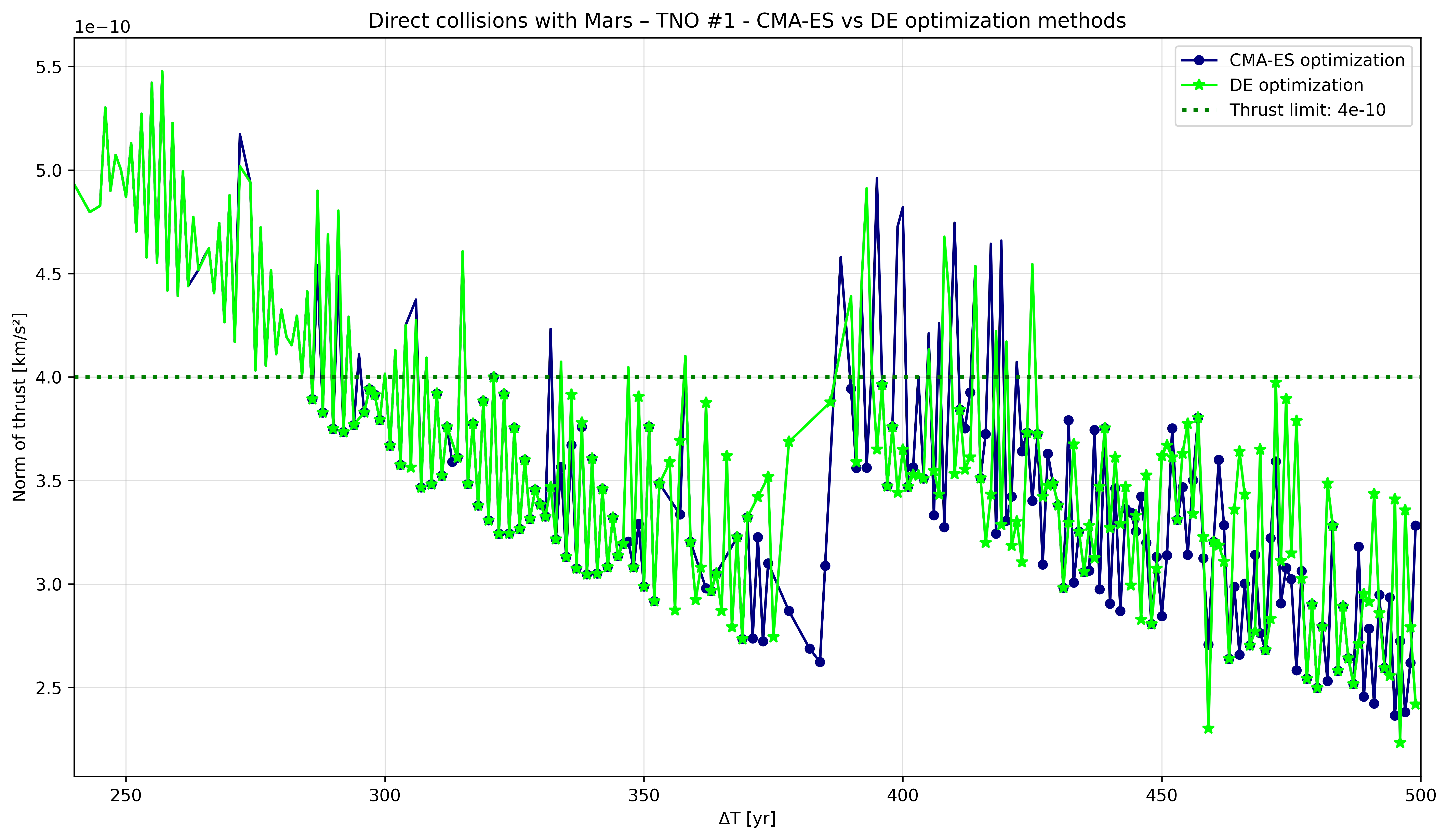}
  \caption{Continuous thrust norm as a function of transfer times $\Delta T$ for direct-impact trajectories of test body TB~\#1. Solid blue lines represent solutions obtained using CMA-ES optimization, and solid green lines represent Differential Evolution (DE) solutions. The horizontal dark green dotted line denotes the imposed thrust limit of $4\times10^{-10}~\mathrm{km\,s^{-2}}$. The collisional gap reflects a temporary loss of admissible impact geometry. Beyond this interval, a modest increase in thrust is required to restore the trajectory to a collision-permitting configuration.}
  \label{fig:4}
\end{figure*}

The corresponding thrust-norm time profiles on Fig.~\ref{fig:4} remained below the imposed limit and showed a characteristic decline in magnitude as the orbit approached the collisional regime.

Representative minima include (only the TB~\#1 solution is shown on Fig.~\ref{fig:3}):
\begin{align}
\text{TB~\#1:}\quad &\Delta V_{\mathrm{opt}} = 3.18~\mathrm{km\,s^{-1}} ~ at~ \Delta T = 384~\mathrm{yr},\\
\text{TB~\#2:}\quad &\Delta V_{\mathrm{opt}} = 2.50~\mathrm{km\,s^{-1}} ~ at~ \Delta T = 480~\mathrm{yr}.
\end{align}

The optimization procedure yielded significant improvements over the spiral transfer, reducing both the mission duration and the propellant requirements. This demonstrates the efficiency gains which is achievable through trajectory optimization for continuous thrust transfers.

\subsection{Gravity-Assisted Transfers}

To evaluate whether a single Neptune encounter can reduce the control effort required for Mars-impacting trajectories, we performed the assessment by analysing a set of flyby configurations defined on a discrete grid specifying both the encounter-plane (b-plane) orientation angle and the distance of closest approach. The azimuth angle $\varphi$ in the b-plane was varied in $5^\circ$ increments and the impact parameter $b$ was scanned from $18$ to $25$ of Neptune radii. At these distances, the gravitational influence of the ring system can be neglected, and the flyby was modelled as a restricted three-body problem. Such a two-dimensional search is computationally expensive, as each candidate geometry requires full pre- and post-encounter propagation under continuous thrust.

To avoid propagating post-encounter trajectories that in general can have lower probability to achieve low 
$\Delta V$ values, we applied a dynamical feasibility filter based solely on a keplerian orbit after the flyby. Only
 those encounters for which the post-encounter orbit satisfied either
\[
|q_{\mathrm{post}} - a_{\mathrm{Mars}}| < 0.5~\mathrm{AU}
\quad \text{or} \quad
\mathrm{MOID} < 0.15~\mathrm{AU}
\]
were propagated with continuous thrust after gravity assist. 

Within the filtered set, only a few encounter configurations produced Mars-impacting trajectories with total control effort comparable to or lower than that of the direct optimized transfers. For test body TB~\#1, a favorable encounter minimum values found:
\[
\Delta V_{\mathrm{tot}} = 3.11~\mathrm{km\,s^{-1}}
\quad \text{at} \quad
\Delta T_{\mathrm{tot}} = 421~\mathrm{yr}.
\]

And similarly for TB \#2:
\[
\Delta V_{\mathrm{tot}} = 2.19~\mathrm{km\,s^{-1}}
\quad \text{at} \quad
\Delta T_{\mathrm{tot}} = 540~\mathrm{yr}.
\]

This indicates that even a single planetary flyby can reduce the energy required to achieve a Mars impact. The magnitude of this benefit depends on the initial orbital geometry: for TB~\#2 the effect is more pronounced due to its initial higher orbital eccentricity. The increased inclination of this orbit does not compensate for the advantage.

\begin{figure*}[htbp]
  \centering
  \includegraphics[width=0.95\textwidth]{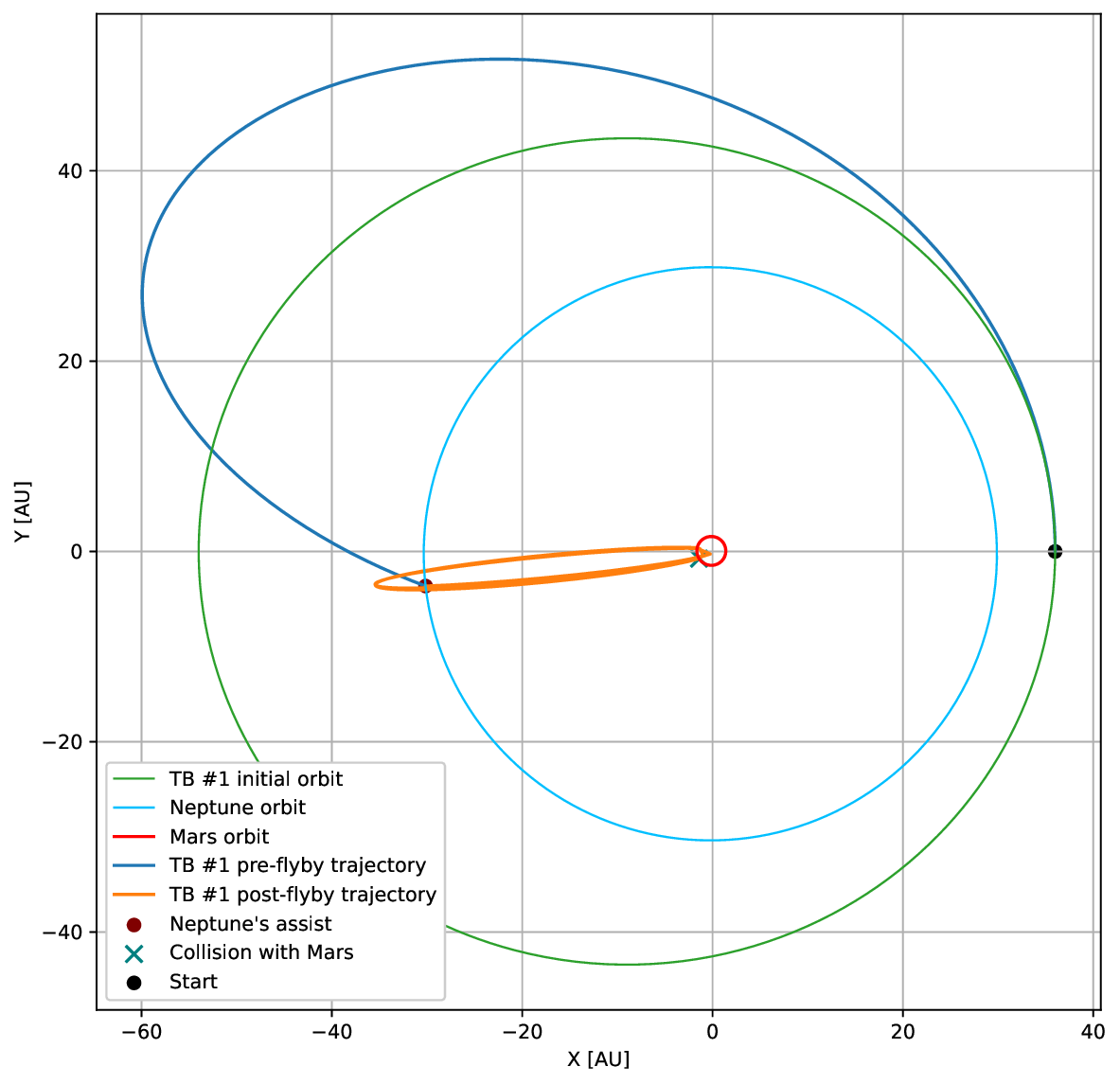}
 \caption{Heliocentric trajectory of test body TB~\#1 projected onto the XY plane. The plot shows the continuous thrust transfer to Neptune (navy blue), the gravity assist (maroon dot), and the resulting highly eccentric post-flyby trajectories (orange). The collision with Mars (teal cross) occurs after the second perihelion passage of the planet.}
  \label{fig:5A}
\end{figure*}

\begin{figure*}[htbp]
  \centering
  \includegraphics[width=0.95\textwidth]{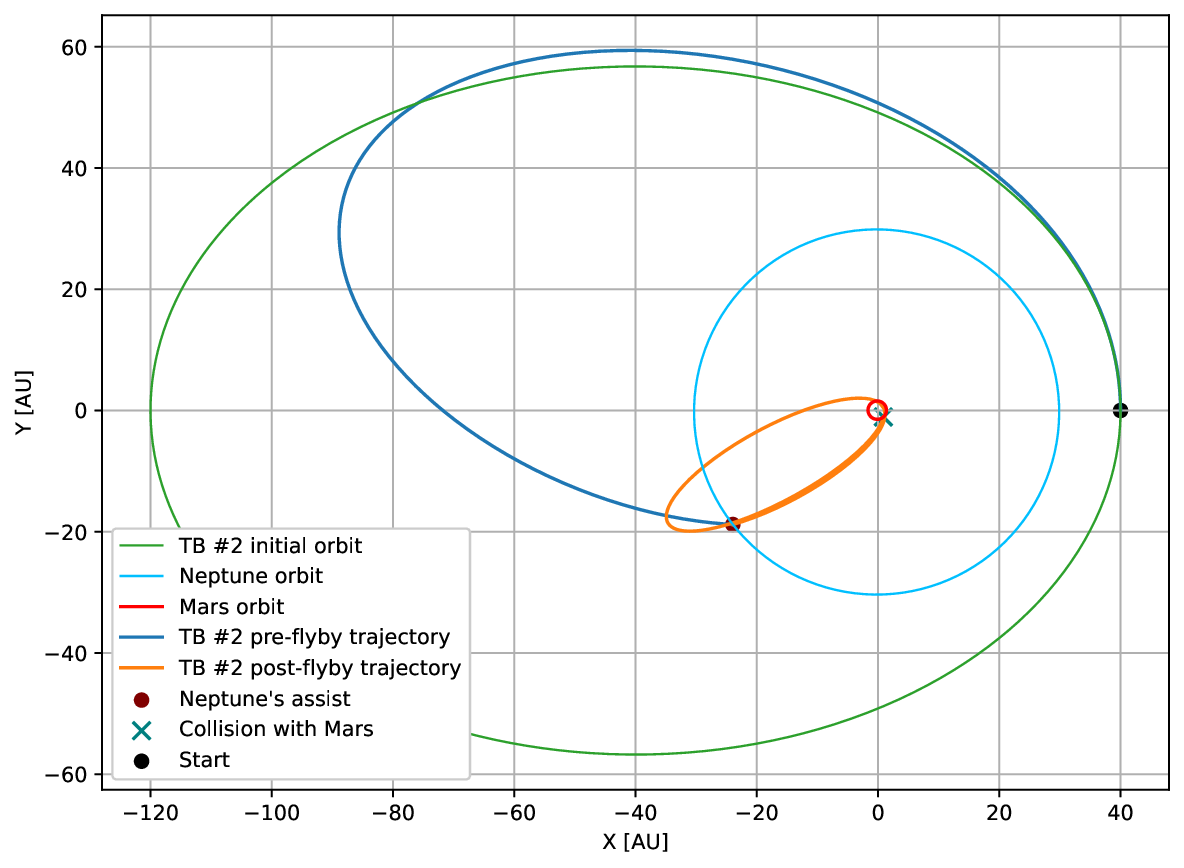}
\caption{Heliocentric trajectory of test body TB~\#2 projected onto the ecliptic XY plane. In this configuration, the post-flyby trajectories (orange) maintain a moderate perihelion distance ($q \approx 1\,\mathrm{AU}$), leading to a collision with Mars (teal cross) after its first perihelion passage.}
  \label{fig:5B}
\end{figure*}

However, the two test bodies exhibit qualitatively different post-encounter dynamics, as illustrated in Figures~\ref{fig:5A} and~\ref{fig:5B}. For TB~\#1, the Neptune flyby produces a highly eccentric orbit with perihelion at $q \simeq 0.032\,\mathrm{AU}$, well within the region typically associated with the extended solar corona. The Mars impact can occur only after two complete post-encounter perihelion passages, during which the test body would experience extreme thermal and radiation conditions. Such a trajectory, while energetically favorable, is physically unsuitable for transporting volatile-rich bodies: at such small heliocentric distances, the object would experience catastrophic mass loss through sublimation and could undergo structural disintegration due to thermal stress. Additionally, non-gravitational forces from outgassing would perturb the orbit, requiring costly trajectory corrections to achieve Mars impact and eliminating much of the $\Delta V$ savings.

TB~\#2, however, follows a significantly more favorable trajectory after the Neptune encounter. The post-flyby orbit has a perihelion distance of $q \approx 1\,\mathrm{AU}$, that slowly decreases under continuous thrust. The Mars collision occurs after the first post-encounter perihelion passage. This geometry represents a dynamically efficient and physically possible path for delivering volatiles to Mars - the post-encounter orbit should provide body's integrity during the perihelion passage. 

These results align with past research indicating that gravitational encounters are a key driver of TNO delivery efficiency toward terrestrial planet-crossing orbits \citep{KaibQuinn2009}.

Table~\ref{Tab1} summarizes the key parameters across all three trajectory types examined in this study. 

\begin{table}[htbp]
\centering
\caption{Summary of the transfer characteristics obtained for the different trajectory classes. Spiral transfer values represent descent to Mars semi-major axis only, not actual collision trajectory. TB \#2 reached an eccentricity threshold at $e = 0.999$ before the object could approach the orbital distance of Mars. For the gravity-assisted solutions, the table reports the $\Delta V_{tot}$ obtained as the sum of $\Delta V_1$ and $\Delta V_2$, the total transfer time as $\Delta T_1 + \Delta T_2$, and the net effective thrust as the combined contribution from the pre- and post-assist phases.}
\resizebox{\linewidth}{!}{
\begin{tabular}{lcccccc}
\hline
\hline
 & \multicolumn{2}{c}{Spiral (retrograde)} & \multicolumn{2}{c}{Direct optimized} & \multicolumn{2}{c}{Gravity-assisted} \\
\cline{2-3} \cline{4-5} \cline{6-7}
Parameter & TB~\#1 & TB~\#2 & TB~\#1 & TB~\#2 & TB~\#1 & TB~\#2 \\
\hline
$\Delta V$ [km\,s$^{-1}$] & $22.785$ & --- & $3.179$ & $2.501$ & $3.107$ & $2.192$ \\
$\Delta T$ [yr] & $1805.008$ & --- & $384.0$ & $480.0$ & $421.0$ & $540.0$ \\
Thrust [km\,s$^{-2}$] & $4 \times 10^{-10}$ & $4 \times 10^{-10}$ & $2.623 \times 10^{-10}$ & $1.651 \times 10^{-10} $ & $4.126 \times 10^{-10}$ & $2.107 \times 10^{-10}$ \\
\hline
\hline
\end{tabular}
}
  \label{Tab1}
\end{table}

\section{Conclusions}

This study assessed the dynamical feasibility of redirecting small volatile-rich 
trans-Neptunian objects onto Mars-impacting trajectories using continuous thrust. 
The adopted two-body model enables the determination of the minimum $\Delta V$ and transfer time 
required for an actual collision with Mars across two of the three investigated trajectory classes. 
The spiral retrograde transfer is the most demanding in both metrics, requiring nearly an order of 
magnitude more $\Delta V$ applied over millennia, rendering this mechanism impractical for any 
realistic Mars-terraforming scenario. Moreover, spiral trajectories cannot efficiently accommodate bodies on orbits with higher initial eccentricities, unless additional control is imposed to regulate 
further eccentricity growth.

By contrast, optimized thrust-direction steering yields direct impact trajectories 
with $\Delta V$ in the range $2.5$ - $3.2~\mathrm{km\,s^{-1}}$ and transfer times of 
$380$ - $540$~yr, with the specific outcomes governed by the geometry of 
the initial orbit. For these cases the required continuous 
thrust remains below the assumed maximum available acceleration of 
$4\times10^{-10}~\mathrm{km\,s^{-2}}$, demonstrating that such transfers fall 
entirely within the dynamical capabilities permitted by the adopted continuous thrust model.

The transfer energetics are strongly influenced by the initial orbital geometry. 
Bodies beginning on more eccentric orbits require significantly lower $\Delta V$ 
effort to reach a Mars-impacting configuration, even when their inclinations are 
higher. For more eccentric orbits, small velocity adjustments made while the object 
is still close to aphelion can produce larger changes in the orbital trajectory, 
enabling impacts at lower $\Delta V$.

Hybrid trajectories that employ a single gravity assist at Neptune can, under favourable configurations, achieve lower total $\Delta V$ than optimized direct transfers. For both test bodies analysed, the most efficient gravity-assisted solutions reduce $\Delta V_{\mathrm{tot}}$ comparing to the direct-transfer baseline. The magnitude of this improvement depends on the initial orbital eccentricity: the more eccentric TB~\#2 benefits most strongly, exhibiting a $\sim12\%$ reduction in $\Delta V_{\mathrm{tot}}$ accompanied by an increase of approximately $12.5\%$ in transfer time.

In these advantageous encounter geometries, the gravity assist reshapes the heliocentric trajectory such that the post-encounter perihelion lies naturally closer to $a_{\mathrm{Mars}}$. As a result, the pre-encounter manoeuvre $\Delta V_{1}$ remains modest, and the subsequent post-encounter correction $\Delta V_{2}$ required to place the body onto a Mars-impacting trajectory becomes significantly smaller, typically by an order of magnitude compared with the pre-assist contribution.

Throughout these solutions, each continuous thrust component in the pre- and post-encounter phases remains individually below the imposed upper bound of $4\times10^{-10}\,\mathrm{km\,s^{-2}}$. Their combined value, however, may exceed this limit in some cases (see the value for TB~\#1), but this aggregate quantity serves only as a comparative metric and does not correspond to any physical acceleration acting along the trajectory. As summarised in Table~\ref{Tab1}, the gravity assist architecture therefore provides a genuine reduction in the total control effort value $\Delta V$, at the price of larger cumulative thrust over time and longer transfer durations.

The dynamical implications of the post-assist state require careful assessment. TB~\#1 attains a perihelion deep inside the solar corona ($q \sim 0.032$~AU). At such extreme distances, non-gravitational perturbations - including sublimation-driven outgassing, solar-radiation pressure, and thermal re-radiation forces such as the Yarkovsky effect - will strongly alter the orbital evolution, potentially leading to substantial mass loss including fragmentation, and an  increase in the required $\Delta V$. Trajectories involving such close solar approaches are therefore unsuitable and should be excluded from further consideration.

Overall, the analysis demonstrates that controlled redirection of trans-Neptunian objects onto Mars-impacting orbits is dynamically achievable under very low continuous acceleration, with transfer times of 380--550~years for bodies originating from favourable regions of orbital phase space, particularly those with moderate to high eccentricities. Orbital inclination plays only a secondary role and does not significantly influence the required $\Delta V$.

Further reductions in $\Delta V$ would likely require sequences of multiple planetary 
assists (e.g., Neptune - Uranus - Saturn chains), but identifying such multi-encounter 
pathways through high-dimensional search spaces is computationally challenging. In 
such architectures, the dominant contributions to the total $\Delta V$ arise from 
the initial segment: from the departure orbit to the first close approach with 
Neptune, and from the terminal segment linking the final assist to the 
Mars-impacting trajectory. Our calculations indicate that the $\Delta V$ for the 
initial segment can be kept below $1~\mathrm{km\,s^{-1}}$ for suitably chosen 
trans-Neptunian starting conditions.

The computational framework will be further extended to incorporate a more realistic 
treatment of physical effects acting on transferred bodies, as well as improved 
selection and optimization strategies for identifying admissible trajectories. 
The methods and algorithms developed here for the purposes of large-scale mass 
transport also have broader applicability in astronautics and in general in studies of 
celestial dynamics. \\

\end{document}